# Characteristics of Interlayer Tunneling Field Effect Transistors Computed by a "DFT-Bardeen" Method

Jun Li,[1] Yifan Nie,[2] Kyeongjae Cho,[2] and Randall M. Feenstra[1,*]

[1]Dept. Physics, Carnegie Mellon University, Pittsburgh, Pennsylvania, U.S.A.
[2]Dept. Materials Science and Engineering, The University of Texas at Dallas, Dallas, Texas, U.S.A.

**Abstract**—Theoretical predictions are made for the current-voltage characteristics of two-dimensional heterojunction interlayer tunneling field-effect transistors (Thin-TFETs), focusing on the magnitude of the current that is achievable in such devices. A theory based on the Bardeen tunneling method is employed, using wavefunctions from first-principles density-functional theory. This method permits convenient incorporation of differing materials into the source and drain electrodes, i.e. with different crystal structures, lattice constants, and/or band structures. Large variations in the tunnel currents are found, depending on the particular two-dimensional materials used for the source and drain electrodes. Tunneling between states derived from the center (Γ-point) of the Brillouin zone (BZ) is found, in general, to lead to larger current than for zone-edge (e.g. K-point) states. Differences, as large as an order of magnitude, between the present results and various prior predictions are discussed. Predicted values for the tunneling currents, including subthreshold swing, are compared with benchmark values for low-power digital applications. Contact resistance is considered and its effect on the tunneling currents is demonstrated.

I. INTRODUCTION

Owing to their very low off-state currents, and steep subthreshold swing when approaching the on state, tunneling field-effect transistors (TFETs) are very attractive devices for low-power electronic applications.[1] In recent years, two-dimensional (2D) layered materials have been studied both theoretically and experimentally for such devices.[2-7] We focus in this work on vertical, interlayer devices in which the tunneling occurs between 2D layers (rather than within a layer). Such devices consist of two electrodes (source and drain), surrounded by one or two gates, as pictured in Fig. 1(a). The source and drain may be separated by one or more layers of insulating material forming the tunnel barrier (such as hexagonal boron nitride, h-BN), or in principle the tunneling can occur simply between the van der Waals (vdW) gap that separates the source and drain. Following Li et al.,[2,3] we refer to such devices as two-dimensional heterojunction interlayer tunneling field-effect transistors (Thin-TFETs).

There are two fundamentally different modes of operation for a Thin-TFET: tunneling between unlike bands, or tunneling between like bands, as schematically illustrated in Figs. 1(b) and (c). For unlike-band tunneling, electrons flow from the valence band (VB) of one electrode to the conduction band (CB) of the other. This is the usual mode for TFETs, providing a steep turn-on of the current when the bands overlap (this mode is also known as band-to-band, Zener, or reverse-bias tunneling). In contrast, for like-band tunneling the electrons flow from VB to VB or CB to CB, i.e. depending on the Fermi-level positions in the source and drain. For the case of 2D materials in particular, this mode yields negative differential resistance (NDR) due to the phenomenon of "lateral momentum conservation" during the tunneling (hence, this mode is sometimes referred to as 2D-2D tunneling).[8-14]

---
[*] feenstra@cmu.edu

In this work, we consider unlike-band tunneling in Thin-TFETs, focusing on the magnitudes of the currents that are attainable in such devices. We employ the Bardeen tunneling approach,[15,16] with wavefunctions from density-functional theory (DFT). The tunneling currents that we obtain differ, by as much as an order of magnitude, from those obtained in some prior theoretical approaches to this problem.[2,3,7-11,13,14] The main goal of this work is to obtain reliable estimates of the magnitude of the tunnel current, for comparison with benchmark values that are needed for low-power digital applications. We discuss the differences between the details of the various theoretical approaches, and argue that our present method provides reasonably reliable estimates for the magnitude of the current (while at the same time recognizing that certain aspects of the problem are not well treated in the present computations).

## II. THEORETICAL METHOD

We employ the Bardeen method for tunneling,[15-17] as described in detail in our previous work which dealt with graphene-based devices.[8] This method is a first-order perturbative approach, which does not permit inclusion of interactions between the electrodes (other than those that produce tunneling). Rather, it treats the electronic structure each electrode *in the absence* of the other, and hence electrodes of differing materials can be easily handled (i.e. without explicitly considering the large unit cells that constitute an epitaxial match between the two materials). However, when the tunneling barrier consists simply of a vdW gap between source and drain (i.e. with no h-BN or other insulator in the barrier) then certainly the interactions between electrodes will not be negligible. Nevertheless, the goal of our work is to evaluate how the magnitude of the tunnel current will vary depending on the material used for the source and drain electrodes, i.e. depending on the overlap of the wavefunctions between the two electrodes. In this regard our computations employing the Bardeen method provide useful information, since we find orders-of-magnitude variations in the tunneling current depending on the materials. We also note that even though the tunnel barriers formed in the vdW gap between 2D electrodes are relatively small, we find that the Bardeen method still works fairly well (accuracy of a factor of 2 – 3, with the currents being *underestimated* by this amount) for the cases we consider, as demonstrated in the Appendix.

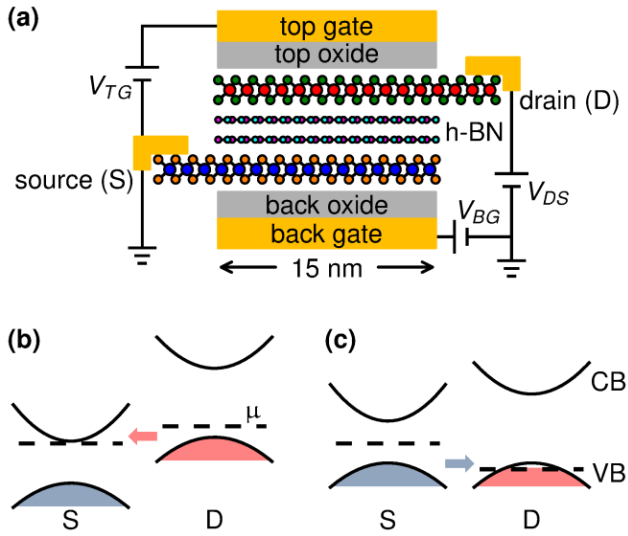

**Fig**. **1**. (a) Schematic view of Thin-TFET. Source and drain electrodes are typically made of transition metal dichalcogenide monolayers, with zero, one or more layers of h-BN as the tunneling barrier. (b) and (c) Unlike-band and like-band tunneling, respectively, showing the valence band (VB) and conduction band (CB) of the source (S) and drain (D) electrodes. Dashed lines represent the Fermi levels of the electrodes. A source/drain overlap length of 15 nm is assumed, as indicated.

In contrast to prior work which employed only a very approximate form of the wavefunction (i.e. just a single plane-wave, SPW),[8,9] in the present work we employ the full form of the wavefunctions as given by the Vienna Ab Initio Simulation Package (VASP),[18] with the projector-augmented wave method.[19] The Perdew-Burke-Ernzerhof form of the generalized-gradient approximation (GGA) for the density functional is used.[20] The wavefunctions are expanded in plane waves with a kinetic energy cutoff of 500 eV, and the convergence criterion for the electronic relaxation is $10^{-4}$ eV. The computation of states over the Brillouin zone (BZ) needed to compute the tunnel current is performed typically with a 32×32×1 or 40×40×1 Monkhorst-Pack k-point mesh.[21] The structure of the transition metal dichalcogenide (TMD) monolayers follows a prior theoretical study.[22] The TFET electrodes are modeled simply as single monolayers (MLs) of the 2D materials (e.g. adjacent planes of Se-W-Se for the case of $WSe_2$, where we refer to that assembly as a ML). Each supercell includes a vacuum region with width of about 20 Å, to minimize the interaction between adjacent supercells.

Energy bands for monolayer (ML) $WSe_2$ and ML $SnSe_2$ are pictured in Figs. 2(a) and (b), respectively, using their hexagonal Brillouin zones (BZs). $WSe_2$ has its VB maximum at the K-point, and it has two CB minima, one at the K-point and the other at a slightly lower energy located at a Q-point between Γ and K. $SnSe_2$ has its VB maximum located between Γ and M, and its CB minimum at the M-point. Energy bands for phosphorene are pictured in Fig. 2(c).[23] The BZ is rectangular in this case, with CB minimum at the Γ-point and a VB maximum that is relatively broad and extends from the Γ-point to a point between Γ and X.

The wavefunctions that we employ from VASP take the form of plane-wave expansions,

$$\psi_{\nu,\mathbf{k}}(\mathbf{r}) = \sum_{\mathbf{G}} \frac{C_{\nu,\mathbf{k},\mathbf{G}}}{\sqrt{V_C}} e^{i(\mathbf{k}+\mathbf{G})\cdot\mathbf{r}} \quad (1)$$

where $C_{\nu,\mathbf{k},\mathbf{G}}$ are the expansion coefficients for band $\nu$, wavevector $\mathbf{k} \equiv (k_x, k_y, k_z)$, and reciprocal lattice vector of the simulation cell $\mathbf{G} = (G_x, G_y, G_z)$. $V_C$ is the volume of the cell, to which the VASP wavefunctions are normalized. We find it more convenient to employ an area $A = L^2$ for the lateral part of the wavefunction, rather than the area of the unit cell $A_C$, and hence we multiply the wavefunctions by a factor of $\sqrt{A_C/A}$. Including a factor of 2 for spin degeneracy, tunnel currents are obtained from[15,16]

$$I = \frac{4\pi e}{\hbar} \sum_{\alpha,\beta} |M_{\alpha\beta}|^2 [f_L(E_\alpha) - f_R(E_\beta)] \delta(E_\alpha - E_\beta) \qquad (2)$$

with matrix element given by

$$M_{\alpha\beta} = \frac{\hbar^2}{2m} \int dS \left( \psi_\alpha^* \frac{d\psi_\beta}{dz} - \psi_\beta \frac{d\psi_\alpha^*}{dz} \right) \qquad (3)$$

where $\alpha \equiv (\mathbf{k}_\alpha, \nu_\alpha)$ and $\beta \equiv (\mathbf{k}_\beta, \nu_\beta)$ label the states of the two electrodes, having energies $E_\alpha$ and $E_\beta$, respectively, and where $m$ is the free electron mass. We choose the normalization length in the $z$ direction to be the supercell period, in which case we need only include the standing wave states with $k_z = 0$ in this computation of the current; thus we henceforth take $\mathbf{k} \equiv (k_x, k_y)$ for both electrodes. In Eq. (2), $f_\alpha$ and $f_\beta$ are Fermi occupation factor for the electrodes, $f_\alpha(E) = \{1 + \exp[(E - \mu_\alpha)/k_B T]\}^{-1}$ and $f_\beta(E) = \{1 + \exp[(E - \mu_\beta)/k_B T]\}^{-1}$, where $\mu_\alpha$ and $\mu_\beta$ are the chemical potentials in the two electrodes, $\mu_\alpha - \mu_\beta = -eV$, where $V$ is the applied bias on the α-electrode relative to the β-electrode.

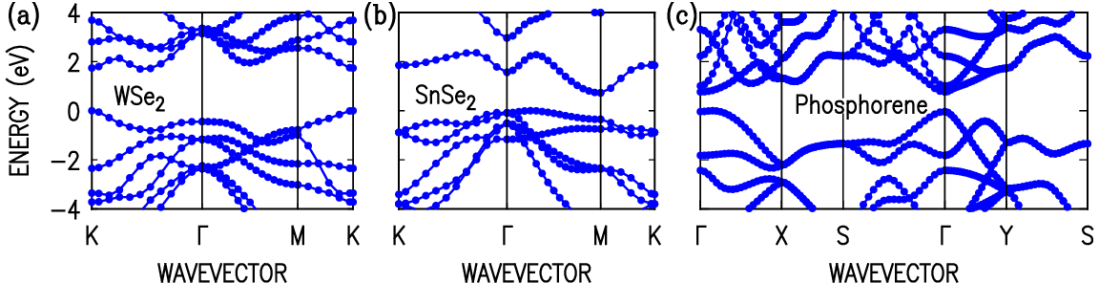

**Fig**. 2. Energy bands for (a) ML WSe$_2$, (b) ML SnSe$_2$, and (c) Phosphorene from DFT computations. For the plots, the zero energy level has been taken to be that of VB maximum for each of the materials.

Utilizing Eq. (1) for the wavefunctions, we evaluate the integrand of Eq. (3) to be

$$\frac{i}{A}\sqrt{\frac{A_{C,\alpha}\,A_{C,\beta}}{V_{C,\alpha}\,V_{C,\beta}}}\sum_{\mathbf{G}_\alpha,\mathbf{G}_\beta}(G_{z,\beta}+G_{z,\alpha})C^{*}_{\mathbf{G}_\alpha,\mathbf{k}_\alpha,\nu_\alpha}$$
$$C_{\mathbf{G}_\beta,\mathbf{k}_\beta,\nu_\beta}\,e^{i(G_{z,\beta}-G_{z,\alpha})z}\,e^{i(\mathbf{k}_\beta+\mathbf{G}_{\beta,\rho}-\mathbf{k}_\alpha-\mathbf{G}_{\alpha,\rho})\cdot\boldsymbol{\rho}} \qquad (4)$$

with $\boldsymbol{\rho}\equiv(x,y)$ and where $\mathbf{G}_\rho\equiv(G_x,G_y)$ for both electrodes. The integrand is evaluated at the point $z$, which is half of the barrier width from the atomic planes of the 2D layers of each electrode. The barrier width is determined by the experimental separation between the electrode materials, if known, or if not then by the average of the layer-separations of the individual electrode materials in bulk form. Considering now the surface integral over the plane separating the electrodes, the only term in Eq. (4) that has any $(x,y)$ dependence is the final exponential term. We have argued previously that a useful model for evaluating the surface integral is to consider a phase coherent area for the wavefunctions in the respective electrodes, given by $A=L^2$ where $L$ is denoted as the coherence length (we have utilized the same $L$ above for the wavefunction normalization). The surface integral of the final exponential in Eq. (4) is then easily evaluated, yielding

$$L^2\,\mathrm{sinc}\!\left(\frac{Lq_x}{2}\right)\mathrm{sinc}\!\left(\frac{Lq_y}{2}\right) \qquad (5)$$

where $q_x\equiv k_{\beta,x}+G_{\beta,x}-k_{\alpha,x}-G_{\alpha,x}$, $q_y\equiv k_{\beta,y}+G_{\beta,y}-k_{\alpha,y}-G_{\alpha,y}$, and $\mathrm{sinc}(u)\equiv\sin(u)/u$. As previously discussed,[11] a somewhat better model is to utilize a *distribution* of phase coherence lengths, in which case the expression of Eq. (5) can be replaced by

$$\frac{L^2}{[1+(q/q_c)^2]^{3/2}} \qquad (6)$$

with $q_c\equiv 2\pi/L$, $q=|\mathbf{q}|=\sqrt{q_x^2+q_y^2}$.

Combining Eqs. (2), (3), (4), and (6), a formal expression for the tunnel current is easily obtained. However, this expression still contains the energy δ-functions of Eq. (2). One could simply broaden those δ-function (e.g. as done in Ref. [7]), but since we are interested in evaluating the steepness of the turn-on for unlike band tunneling, we wish to avoid such broadening. Hence we convert the sum in Eq. (2) for one of the electrodes, say the β-electrode, into an integral over $\mathbf{k}_\beta$ (two-dimensional wavevector), and then into an integral over energy according to

$$\sum_{\mathbf{k}}=\frac{A}{(2\pi)^2}\iint d^2k=\frac{A}{(2\pi)^2}\int dE\int\frac{d\ell_k}{|\nabla_k E|} \qquad (7)$$

where $\int d\ell_k$ is a line integral in **k**-space along a constant-energy contour. In this way the energy δ-function can be used to evaluate the energy integral, with the line integral and the gradient term evaluated at the specific energy of the state of the α-electrode. Hence, the expression for the tunnel current becomes

$$I = \frac{e}{\pi\hbar} \frac{A_{C,\alpha} A_{C,\beta} A}{V_{C,\alpha} V_{C,\beta}} \sum_{\mathbf{k}_\alpha, \nu_\alpha, \nu_\beta} \left[ f_\alpha(E_{\mathbf{k}_\alpha,\nu_\alpha}) - f_\beta(E_{\mathbf{k}_\alpha,\nu_\alpha}) \right] \int \frac{d\ell_{k_\beta}}{|\nabla_k E_\beta|} |M_{\alpha\beta}|^2 \quad (8)$$

where

$$M_{\alpha\beta} = \frac{\hbar^2}{2m} \sum_{\mathbf{G}_\alpha, \mathbf{G}_\beta} \frac{(G_{z,\beta} + G_{z,\alpha}) C^*_{\mathbf{G}_\alpha,\mathbf{k}_\alpha,\nu_\alpha} C_{\mathbf{G}_\beta,\mathbf{k}_\beta,\nu_\beta} e^{i(G_{z,\beta}+G_{z,\alpha})z}}{[1+(q/q_c)^2]^{3/2}} \quad (9)$$

with $f_\alpha$ and $f_\beta$ defined following Eq. (3), $q$ and $q_c$ defined following Eq. (6), and where, again, the line integral along the constant-energy contour in the β-electrode is evaluated at the specific energy of each state of the α-electrode. Equation (8) as written provides the current over an $L \times L$ area, so that current density is given by $I/A$. For our results in the following section, we consider an overlap length between source and drain of 15 nm, so the current per unit electrode width is $I/A$ times 15 nm.

To evaluate the β-electrode line integrals, a small area around each k-point that is exclusive to that point is defined (i.e. a "mini BZ", with same shape as the BZ but smaller in area by a factor of $n^2$ for a $n \times n$ mesh). This area is split up into a series of triangles utilizing the lines joining the particular k-point with its neighbors. Using linear interpolation of the energies between neighboring k-points, the constant-energy contour is defined within each triangle and hence in a piecewise linear fashion across the entire BZ, and similarly the magnitude of the gradient $|\nabla_k E|$ is evaluated along the contour. The magnitude of $|M_{\alpha\beta}|^2$ varies along the line integral, in accordance with both the $q$ values and the values of the $C_{\mathbf{G}_\beta,\mathbf{k}_\beta,\nu_\beta}$ coefficients; the former are accurately known, and the latter are evaluated depending on which particular k-point is nearest the specific point on the contour. We perform linear interpolation of the $|M_{\alpha\beta}|^2$ values from neighboring k-points when we evaluate the line integral.

For these evaluations, in the β-electrode, all states must be fully defined across the entire BZ (i.e. not just the irreducible wedge of the BZ). For this purpose, we transform both the energies (as scalar quantities) and the plane-wave coefficients (as vector quantities) from the irreducible wedge to all other parts of the BZ. For the coefficients, transformation to negative **k** values is accomplished by time reversal, in which the complex conjugates of the wavefunctions are computed, since the real-space unit cells of the materials lack inversion symmetry in some cases. As an example of intermediate quantities in our computation, we display in Fig. 3 the constant-energy contours of the WSe$_2$ conduction band, evaluated at an energy of 0.06 eV above the conduction band minimum at the K-point. The contours are shown in a repeated-zone scheme, with each zone representing one of the $\mathbf{G}_\beta$ terms from the summation of Eq. (9). (Note that for the particular 32×32 mesh used here, there is not a k-point directly at the K corner of the BZ; nevertheless in our computations we add that point from a separate VASP run, e.g. using a 12×12 mesh, to permit a reasonably good description of the contours even very near the K-point).

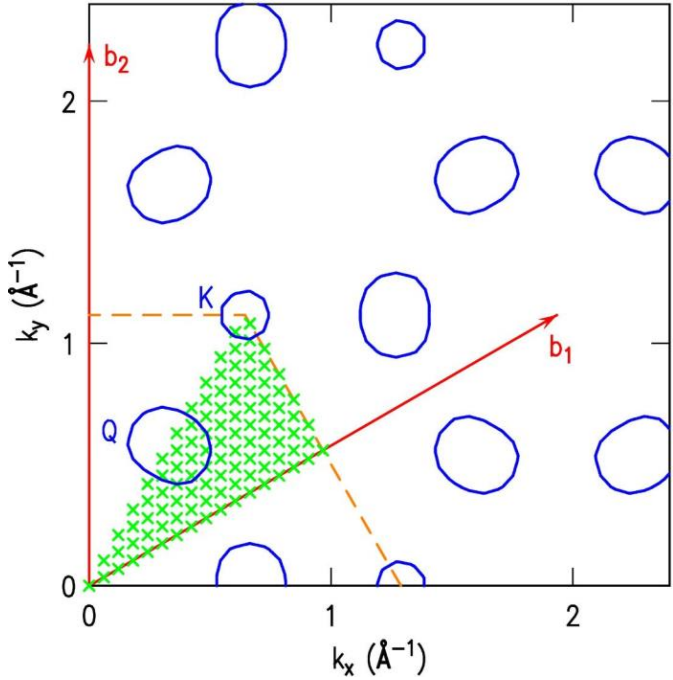

**Fig. 3.** Section of reciprocal space for ML WSe$_2$, showing the k-points (x-marks) in the irreducible wedge of a 32×32 Monkhorst-Pack mesh in the BZ. Orange dashed lines show the edge of the BZ, with reciprocal vectors $\mathbf{b}_1$ and $\mathbf{b}_2$ indicated. Constant-energy contours are represented by blue ovals, for an energy of 0.06 eV above the conduction band minimum at the K-point; contours around the K-points and Q-points are apparent.

Concerning the states of the α-electrode, in principle we should also extend those over the entire BZ in a similar manner as for the β-electrode. However, in many cases (e.g. for the same symmetry of both electrodes and no angular misorientation between them) it suffices to simply multiply the current obtained from a particular state associated with a k-point within the irreducible wedge of the BZ by a suitable factor (e.g. multiplier of 12 for a general point in the irreducible wedge of a hexagonal BZ). Additionally, it is important to realize that since we have normalized the wavefunctions to the volume $AV_C/A_C$ then we have, formally speaking, $A/A_C$ k-points in the sum over $\mathbf{k}_\alpha$ in Eq. (8). Hence, for the $n \times n$ mesh of k-points that we actually use in the computation, the total current must be multiplied by $A/(A_{C,\alpha} n^2)$. As just described, it is clear that the treatment of the two electrodes is quite different in our methodology. Of course, it is possible to swap the sense of the electrodes (together with changing the sign of the applied bias voltage), so that we can choose which electrode to be the α- or the β-one. Generally, it is advantageous to place the electrode with the flattest band as the α-one, so that the spacing of the constant-energy contours in the β-electrode is as small as possible.

To model the electrostatics of the Thin-TFET, we employ the method described in Ref. [2], which solves a one-dimensional (1D) Poisson equation in the z direction. In this model, the difference of Fermi levels of source and drain is determined by the applied source-drain voltage. The source-drain band alignment is determined by that voltage together with the electron affinities and band gaps of the electrodes and the detailed parameters associated with the gates (dielectric constants, work functions, gates voltages). One gate is held at a fixed potential, and a voltage applied to the other gate then acts to tune the band alignment. In this 1D model, the current density is uniform over the 15 nm overlap area of the electrodes (this assumption of uniform current density over the overlap region is consistent with results from other reports,[24,25,26] discussed in more detail below, which include the possibility of in-plane variation in the potentials and current densities over the electrodes). We also mention that our electrostatic computation uses the DFT-generated density of states, which includes multiple bands (important e.g. for the CB of $WSe_2$), whereas the model of Ref. [2] employs only a single-band effective-mass treatment.

III. RESULTS

We focus on results for 2D materials that have band gaps, for which unlike-band tunneling will produce a steep slope at the onset of the current. In particular, we consider chalcogenide materials (i.e. containing S, Se or Te) as well as phosphorene (Phos). For heterojunction devices, an important criterion in choosing the respective materials of the source and drain is the energy offset between the CB edge of one material (material 2) relative to the VB edge of the other (material 1), i.e. $\Delta E_{CV} = E_C^{(2)} - E_V^{(1)}$. Ideally this energy difference will be relatively small, so that the appropriate band edges are approximately aligned (relative to the vacuum level) without application of large gate voltages. One such heterojunction that has been previously proposed in this regard is $WSe_2$-$SnSe_2$, with $\Delta E_{CV} = 0.2$ eV.[3] We examine this case in detail here, and compare it to a Phos-Phos TFET (which does not have a small $\Delta E_{CV}$ value, but nonetheless is interesting for comparison purposes since the resulting tunnel currents are larger).

For all computed current results presented in this work, we consider a coherence length of 10 nm, a source/drain overlap length of 15 nm, equivalent oxide thickness (EOT) in the gate dielectrics of 1 nm for both gates, and drain source bias ($V_{DS}$) of − 0.2 V. A vdW gap, i.e. zero layers of h-BN in the tunnel barrier, is assumed. We also choose different gate work functions (within realistic range) for different devices in order to better align the bands of the source and drain electrodes and thusly maximize the current.

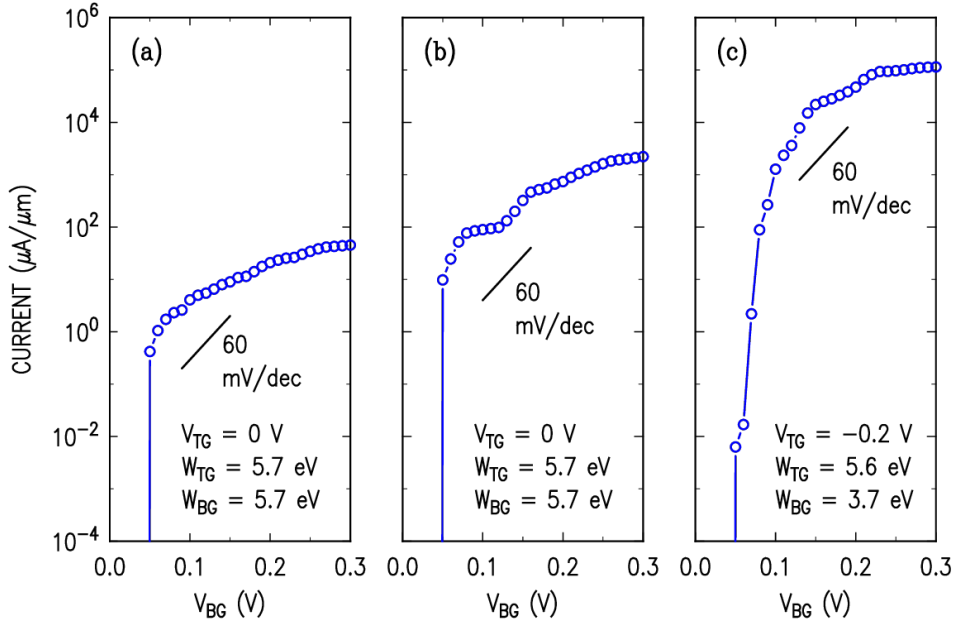

**Fig**. 4. Current flowing from source to drain, $I_D$, as a function of bottom gate voltage, $V_{BG}$, for (a) a WSe$_2$-SnSe$_2$ device with their lattice being aligned (b) a WSe$_2$-SnSe$_2$ device with 30° rotational misalignment between their lattices and (c) a Phos-Phos device with their lattices aligned. Values are listed for top gate voltage ($V_{TG}$) and work functions of top gate ($W_{TG}$) and bottom gate ($W_{BG}$).

Figure 4(a) shows the computed $I_D$ vs. $V_{BG}$ characteristic for a WSe$_2$-SnSe$_2$ TFET, with the lattices of WSe$_2$ and SnSe$_2$ being aligned. The current is caused by tunneling of electrons from the VB maximum of WSe$_2$ to the CB minimum of SnSe$_2$. Note that there is a wavevector mismatch between the tunneling states. Figure 4(b) shows the result still for a WSe$_2$-SnSe$_2$ TEFT, but with 30° rotation between their lattices in order to better align the wavevectors of the SnSe$_2$ CB minimum with those of the WSe$_2$ VB maximum. An obvious increase in the magnitude of the current is observed. Figure 4(c) shows the result for the Phos-Phos device, where we see that the tunneling currents are much larger than those of the WSe$_2$-SnSe$_2$ (either 0° or 30°) device. This difference originates from the overlap matrix elements of the two cases, i.e. Eq. (1), which for WSe$_2$-SnSe$_2$ are relatively small due to the detailed nature (symmetry) of the wavefunctions, whereas for Phos-Phos the values are much larger.

From the $I_D$-$V_{TG}$ characteristic shown in Fig. 4, we can extract values that are useful for benchmarking of the device performance. Specifically, the current at which the subthreshold swing (SS) changes from <60 mV/dec to >60 mV/dec is denoted by $I_{60}$. In addition, an ON current for the device, $I_{ON}$, can be characterized by taking the current at a gate voltage that is +0.2 V greater than the onset voltage of the characteristic. We multiply the current densities from the computations by the overlap length in order to obtain currents per unit width of the device, µA/µm. In Fig. 5 we display these two quantities, $I_{60}$ and $I_{ON}$, for the WSe$_2$-SnSe$_2$ (0° and 30°) and Phos-Phos devices jut discussed as well as for a variety of other Thin-TFETs. Structures of most of the materials in Fig. 5 follow a prior theoretical study.[22] Additionally we considered other 2D materials such as GaSe.[27] Again, we only consider cases where the energy difference $\Delta E_{CV} = E_C^{(2)} - E_V^{(1)}$ is relatively small, with $\Delta E_{CV}$ deduced from the DFT computations (We note that DFT is well known to underestimate experimental band gap values.[28] An approximate correction to the band gaps will cause a right-shift of our results in Fig. 5 by a quarter of the sum of the DFT band gap values of the source and drain electrodes.[22]). On this plot we also include typical desired values for these quantities for low-power digital applications, $I_{60} = 1$ µA/µm and $I_{ON} = 200$ µA/µm.[29,30,31] We see that the WSe$_2$-SnSe$_2$ (30°) device satisfies these benchmark values, while the WSe$_2$-SnSe$_2$ (0°) TFET falls below the desired values.

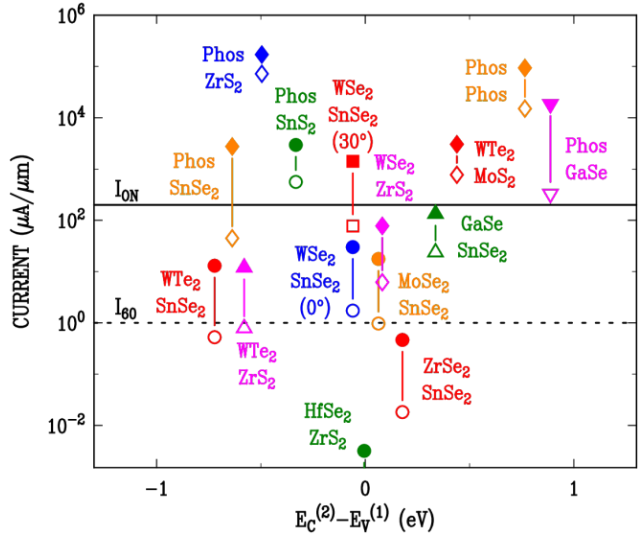

**Fig. 5.** Results for $I_{60}$ (open symbols) and $I_{ON}$ (closed symbols), for Thin-TFET devices made from the materials shown. Electrons flow from the VB of the upper material (denoted 1) to the CB of the lower material (denoted 2), with the difference between these band edges given by $\Delta E_{CV} = E_C^{(2)} - E_V^{(1)}$. The difference $\Delta E_{CV}$ is deduced from the DFT computation. Desired values for $I_{ON}$ and $I_{60}$ are indicated by solid and dashed lines, respectively.

The Phos-Phos device shows a relatively large current, with $I_{60} = 1.50 \times 10^4$ µA/µm and $I_{ON} = 9.39 \times 10^4$ µA/µm. We emphasize that our theory ignores any modifications to the band structure of such a device due to interactions between the electrodes, and indeed these have been shown to be large for the case of Phos-Phos tunneling devices by Constantinescu et al.[7] Those authors argued that a tunnel barrier consisting of one or few layers of h-BN is desirable in order to reduce the interactions between the Phos electrodes. To roughly estimate the inclusion of h-BN in the barrier, we can reduce our current by a factor of 50 for each layer of h-BN added (this factor of 50 arises from explicit computations for graphene/h-BN/graphene tunnel junctions, discussed elsewhere,[32] and is in very good agreement with experimental measurements for this system[33]). For the Phos-Phos TFET, including a single layer of h-BN then yields an ON current $I_{ON} = 1.9 \times 10^3$ µA/µm, still well above the desired values.

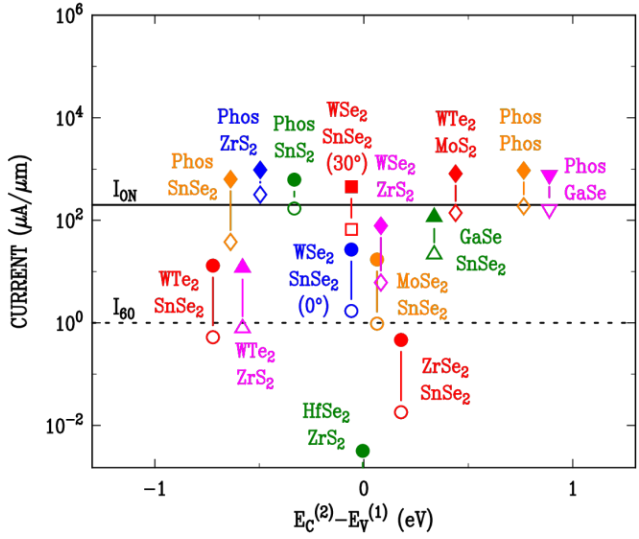

**Fig. 6**. Same caption as for Fig. 5, but now with contact resistance of 80 Ω•µm to both source and drain.

Thus far we have not included any contact resistance in the simulation. To investigate the effect of the contact resistance on tunneling currents, we add contact resistance of 80 Ω·µm to both source and drain based on the recommendation of the 2011 edition of the International Technology Roadmap for Semiconductors (ITRS) and its 2018 node.[34,35] Computations of the tunnel current then proceed iteratively, adjusting the voltage drop across the device based on the current from the prior iteration. The resulting $I_{ON}$ and $I_{60}$ are plotted in Fig. 6. Compared with Fig. 5, it is clear that the tunneling currents with larger magnitude are more affected by the contact resistance, since the maximum possible $I_{ON}$ is now 1250 µA/µm = (0.2 V)/(160 Ω·µm). The ON currents for TFETs such as Phos-Phos, Phos-ZrS$_2$, and WTe$_2$-MoS$_2$ closely approach this limit. For the case of a Phos-Phos device with one layer of h-BN as the tunnel barrier, as discussed in the previous paragraph, we have estimated the ON current to be $1.9 \times 10^3$ µA/µm which is considerably greater than the limiting value of 1250 µA/µm. Hence we expect the ON current of such a device to also exceed the desired value of 200 µA/µm even in presence of the contact resistance.

## IV. DISCUSSION

Generally, Fig. 5 predicts that Thin-TFETs using different 2D materials as electrodes produce $I_{ON}$ and $I_{60}$ values that vary across several orders of magnitude. In particular, devices using Phos as one or both electrodes have currents that are much larger than for devices using TMD as both of the electrodes. There are two reasons for this difference. First, as shown in Fig. 2(c), the VB maximum of Phos is at the Γ-point. States at the Γ point experience a lower tunneling barrier in the vdW gap than do states with finite parallel wavevector **k**, such as the K-point band-edge states in WSe$_2$.

To illustrate this dependence, let us consider tunneling between two electrodes of ML WSe$_2$, as shown in Fig. 7. Figure 7(a) shows the potential (ionic plus Hartree plus exchange-correlation) for both 1 ML and 2 ML WSe$_2$. Within the Bardeen approximation we can consider the two electrodes independently, i.e. using the potential and wavefunctions for the 1ML case and taking the product of the tails of the wavefunctions for the two electrodes at the midpoint of the barrier that separates them. Hence, to understand the magnitude of the tunnel currents it suffices to examine the states of the individual ML electrodes, and in Fig. 7(b) we show two states of ML WSe$_2$, one from the VB maximum at the K point of the BZ and the other from the highest lying VB band at the Γ point (energy of $-0.44$ eV relative to the VB maximum). These wavefunctions are evaluated at a general $(x,y)$ point (i.e. without special symmetry) in the unit cell. As seen in Fig. 7(b), the K-point state has a significantly faster decay in the vacuum than the Γ-point state, due to the nonzero lateral momentum, $|\mathbf{k}| = 1.289$ Å$^{-1}$, of the former state. In general, a state with energy $E$ and nonzero value of $|\mathbf{k}|$ will have a decay constant in the vacuum given by $\kappa = \sqrt{2m(E_{VAC} - E)/\hbar^2 + |\mathbf{k}|^2}$, where $E_{VAC}$ is the vacuum level and $m$ is the free-electron mass.[36,37] Such states decay in the vacuum as if they experience an effective barrier that is *larger* than the nominal one by an amount $\hbar^2|\mathbf{k}|^2/2m$. For the lateral wavevector of 1.289 Å$^{-1}$ we have $\hbar^2|\mathbf{k}|^2/2m = 6.33$ eV, a substantial increase in the effective barrier, and indeed the decay constants deduced from the slopes of the wavefunction tails in Fig. 7(b) are in good agreement with values obtained from this formula for $\kappa$. Thus, states with nonzero lateral momentum experience a much faster decay in the barrier (even for the case of only a van der Waals barrier), and hence the tunnel current that occurs between K-point states will be much less than that which occurs between Γ-point states.

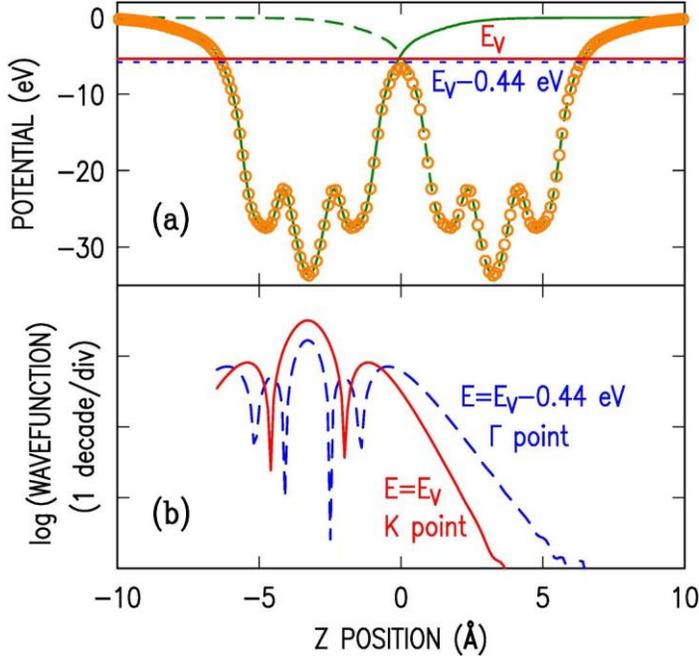

**Fig. 7**. (a) Potential, averaged over $(x,y)$, for 2 ML WSe$_2$ (orange circles) and 1 ML WSe$_2$ (solid and green dashed lines, for two separate MLs). The energy of the VB maximum is indicated, $E_V$ (solid red horizontal line, at 5.38 eV below the vacuum level, which is at 0 eV on the plot), along with energy of a state at $E_V - 0.44$ eV (dashed blue horizontal line). (b) Wavefunctions for 1 ML WSe$_2$ for a K-point state at the VB maximum (solid red line) and a Γ-point state at $E_V - 0.44$ eV (dashed blue line).

The second effect that gives rise to the variation in tunnel current depending on the electrode material, as seen in Fig. 5, has to do with the specific form of the wavefunction that occurs at zone-edge states, producing in certain cases orthogonality (in the overlap matrix element of Eq. (9)) between the states of the source and drain electrodes. This effect can be explained by reference to a two-band nearly free electron model, in which states of the two bands are mixed. According to this model, wavefunctions for states at the BZ edge are very different from those elsewhere in the BZ. An energy gap opens at the BZ edge, with states on either side of the gap having standing wave type wavefunctions of the form $f(z)\cos(\mathbf{k}\cdot\boldsymbol{\rho})$ and $g(z)\sin(\mathbf{k}\cdot\boldsymbol{\rho})$, respectively, where $f(z)$ and $g(z)$ are general functions describing the $z$ part of the wavefunctions. Many TMD monolayers have their VB maximum and/or CB minimum at the K point, which is at the BZ edge. When we consider unlike-band tunneling and compute the overlap matrix element, then because of the orthogonality between the $\cos(\mathbf{k}\cdot\boldsymbol{\rho})$ and $\sin(\mathbf{k}\cdot\boldsymbol{\rho})$ parts of the wavefunctions (i.e. even with the two states being centered at different $z$ values), we find in certain cases a result of zero.

Comparing our results with those of prior theories, we find that there are significant discrepancies between the various treatments. One of these arises from the use of a different form from Eq. (3) for evaluating the matrix element:[7,9]

$$M_{\alpha\beta} = \int d\mathbf{r}\, \psi_\alpha^* V(\mathbf{r}) \psi_\beta \qquad (10)$$

where $V(\mathbf{r})$ is a "scattering potential" of the tunnel barrier, and $\psi_\alpha$, $\psi_\beta$ are wavefunctions of the source and drain electrodes respectively as in Eq. (3). The integral is evaluated over the entire tunnel barrier volume. Equations (3) and (10) will, in general, yield quite different results. There is however some similarity between them, since both equations depend on the difference, $\mathbf{q}$, between the lateral wavevectors for the states in the two electrodes. All authors dealing with 2D tunneling devices evaluate this part of Eqs. (3) or (10) in a similar way, yielding some sort of Gaussian or Lorentzian form that falls off for $|\mathbf{q}|$ values above some critical value (which is inversely proportional to the coherence length $L$). This is the "wavevector conserving" part of the overlap matrix element, and there is general agreement on this part. However, aside from that wavevector-conserving part, the remainder of the matrix element depends on the detailed nature of the wavefunctions as well on the different forms of Eqs. (3) and (10). These portions of the matrix elements from Eq. (3) compared to (10) will, in general, yield quite different results.

For example, a recent report considering a Phos-Phos TFET with one layer of h-BN as the tunnel barrier has employed Eq. (10) for their current computations.[7] In Section III, we have estimated an ON current (with no contact resistance) for such a device to be $1.9\times10^3$ μA/μm. This current is about an order of magnitude smaller than that in Ref. [7]. We believe that this discrepancy arises from the use in that work of Eq.

(10) for the tunneling matrix element (with the potential $V$ taken to be the full potential of the h-BN barrier), whereas we have employed Eq. (3). We emphasize that we are in full agreement with nearly all of the results of Ref. [7], including their voltage-dependence of the current for both like-band and unlike-band tunneling. We have disagreement only on the issue of the absolute magnitude of the tunnel current. (We also note that Ref. [7] explicitly includes the h-BN barrier layer in the wavefunction evaluation, whereas we have only estimated the influence of the h-BN, but it seems unlikely that this difference will lead to a significant *increase* in the current).

Apart from the discrepancy resulting from which form to use for evaluating the tunneling matrix element, a second discrepancy arises from the use of wavefunctions that consist of a single plane-wave (SPW), rather than the full form as in Eq. (1). The SPW wavefunctions lead to a matrix element that, aside from the "wavevector conserving" part discussed above, is essentially independent of the particular states involved in the tunneling. As discussed above in connection with the two-band nearly free electron model, we find in the VASP wavefunctions a large dependence of the matrix elements on the particular states; for all states the magnitude of the matrix element is much different than that obtained from the SPW theory. Theories using essentially the SPW wavefunctions for TMD devices have been employed in several recent works,[2,3,14] employing a constant matrix element value of 0.02 eV (for $|\mathbf{q}| = 0$), obtained by matching to experiment employing an argument involving interlayer charge transfer time.[3] From our DFT results, we find the value of the matrix element for the WSe2-SnSe2 (30°) device to be 0.06 eV, i.e. 3× larger than that in Ref. [3]. Our current (which is proportional to the square of the matrix element) is therefore about an order of magnitude larger, as shown in Fig. 4(b). As revealed by our results of Figs. 4, 5 and 6, the matrix element (and hence the tunnel current) is very dependent on what 2D materials being used for the Thin-TFETs.

Finally, we note that several computations using a theory other than the Bardeen formalism for interlayer TFETs have been reported.[24,25,26] In these works, detailed device simulations are performed using the non-equilibrium Green's function (NEGF) formalism. In Ref. [24], the authors consider a MoS$_2$-WTe$_2$ TFET with 1 nm thick h-BN as the tunnel barrier, with a coupling term between electrodes set by reference to experimental results for graphene/h-BN.[38] To compare with their results, we perform computations using their choice of parameters ($V_{DS}$ = 0.3 V, overlap length of 20 nm, etc.), except with a vdW gap (zero layers of h-BN, corresponding to separation between chalcogen planes of opposing TMD electrodes of about 0.33 nm). We then roughly estimate the current with the h-BN included by dividing the result by 2500, corresponding to two layers of h-BN (0.66 nm) plus our vdW interlayer thickness of 0.33 nm. In this way, we obtain an ON current of 4.0 µA/µm for a coherence length of 5 nm or 10 nm, or of 5.4 µA/µm for the coherence length of 20 nm. These values are in good agreement with those reported in Ref. [24].

In both Refs. [25] and [26], the authors consider a Thin-TFET with only a vdW gap (i.e. no hBN). The first of these reports studies a MoTe$_2$-SnS$_2$ TFET and assumes equal and opposite voltages applied to the two gates. As a result, relatively low tunnel currents are obtained, since the overlap of their VB of MoTe$_2$ with the CB of SnS$_2$ is not maximized (in order to maximize the current at ON state, the overlap of the VB of one electrode with the CB of the other should be maximized subject to constraints by their Fermi levels, i.e. the overlap of tunneling bands should be made close to the difference between Fermi levels of the two electrodes). In Ref. [26], a $MoS_2$-$WTe_2$ TFET is considered and a better electrostatic arrangement was used (with one gate held at fixed potential and a varying voltage applied to the other, i.e. the same as what we employ in the present study). Currents as high as 1000 µA/µm is obtained, for a gate voltage of $V_{TG}$ = − 0.3 V. We have applied our method to their arrangement ($V_{DS}$ = − 0.3 V, overlap length of 30 nm, EOT = 0.5 nm and $V_{BG}$ = 0.5 V), and we find an ON current that is about 6× larger than their result. This difference might arise from the tight-binding approximation used in Ref. [26], or the Bardeen approximation of our work. However, since the electrostatics model used in Ref. [26] is considerably more sophisticated than our 1D model, we feel that further investigation of this aspect of the problem is warranted, in order to better compare the theories.

It is important to remark that the results presented here should be viewed as only approximate estimates of the tunnel currents. The Bardeen method is based on the first-order perturbation theory (hence only requiring knowledge of only the eigenstates of each electrode in the absence of the other), and as such it is a convenient method for obtaining estimates of the tunneling current. However, modifications to the band structure of total system due to interaction between the electrodes are ignored, which is expected to be quite a significant approximation for the case of zero layers of h-BN between the electrodes. Additionally, our computations do not include effects of h-BN interlayers, except for graphene/h-BN/graphene devices where we have included the h-BN.[32] We find in that case the presence of the h-BN (aligned with the graphene) produces about a 50× reduction in the current for each layer of h-BN. However, misalignment of the h-BN and graphene could well produce additional reductions in the current. Similarly, for electrode materials other than graphene, it is possible that reductions to the current (beyond 50× per layer) due to lattice mismatch between the h-BN and the electrodes could well occur.

## V. CONCLUSIONS

The goal of our work is to provide reliable estimates of the magnitude of the tunneling current in Thin-TFETs, which can be benchmarked against values that are appropriate for low-power digital applications. We find a considerable spread in the results depending on the materials used for the electrodes (due to the overlap matrix elements, i.e. considering effects that go beyond those due to wavevector conservation alone). As such, we feel that this work will be useful in choosing among the various materials with which to fabricate Thin-TFETs. The considerations described here regarding the detailed form of the wavefunctions, i.e. their symmetry and momentum-dependent decay constant, would also apply to TFETs made from three-dimensional (3D) materials.[1] However, the Thin-TFETs have the potential advantage that the tunneling can occur across a thin insulating layer (e.g. an h-BN layer, or in principle just a vdW gap), as opposed to the tunneling across the bandgap of a semiconductor depletion layer that is generally utilized in a 3D TFET. This depletion layer requires substantial doping and/or electrostatic gating, such that the distance across it is sufficiently small to enable large current. For the Thin-TFETS considered here, the predicted ON currents are found to be relatively large, resulting from the very thin tunnel barrier (i.e. the vdW gap) that is assumed in the simulations.

## ACKNOWLEDGMENT

This work was supported by the Center for Low Energy Systems Technology (LEAST), one of six centers of STARnet, a Semiconductor Research Corporation program sponsored by Microelectronics Advanced Research Corporation (MARCO) and Defense Advanced Research Projects Agency (DARPA). We thank S. C. de la Barrera, J. J. Nahas, A. C. Seabaugh, and H. G. Xing and for useful discussions.

## APPENDIX

In the main body of this work we consider interlayer tunneling between two 2D electrodes, i.e. with electrodes extending in the (*x,y*) directions and tunneling in the *z* direction through a barrier. An exact solution for this problem using DFT wavefunctions does not exist, and we have employed the Bardeen method to solve this problem. In order to gain some insight into the accuracy of the Bardeen method, we considered a model problem in one dimension (1D), for which both the exact and the Bardeen solutions are readily available.

Our model problem is constructed by employing the plane-averaged potential for WSe$_2$ (as shown in Fig. 7), using the 2-ML potential for our "exact solution" and the 1-ML potentials of two opposing electrodes for the "Bardeen solution". For both situations we truncate the potentials such that they are a constant within the electrodes; specifically, we take the value of the potential at its local minimum that occurs near the Se atoms, and use that for all locations deeper into the electrode. The resulting model potentials are shown in Fig. 8. By taking the potentials to be a constant within the electrodes, we are able to easily obtain both the exact and the Bardeen solutions. (For ease of language we are referring to the locations where the potential is a constant as the "electrode" and elsewhere as the "barrier", although in reality no such division between these two regions is made in our solutions since we employ a completely general solution of the Schrödinger equation in 1D). We consider electrodes of thickness *D* for both problems, as indicated in Fig. 8. We solve the problem for a general *D* value, and then consider small *D* values of 3 – 5 Å, as applicable to actual ML TMD or phosphorene electrodes.

First we consider the exact solution to this one-dimensional problem. For a state with some energy *E* relative to the potential within the electrode, starting with an outgoing plane wave in the right-hand electrode of the form $Ce^{ikz}$ with $k = \sqrt{2mE/\hbar^2}$, we integrate Schrödinger's equation back through the barrier region and into the left-hand electrode.[39] We then normalize the wavefunction over the entire space of the two electrodes plus the barrier, and hence we determine $|C|^2$. The tunneling current associated with this state is then given by $j_z^{\text{exact}} = \hbar e k |C|^2 / m$.

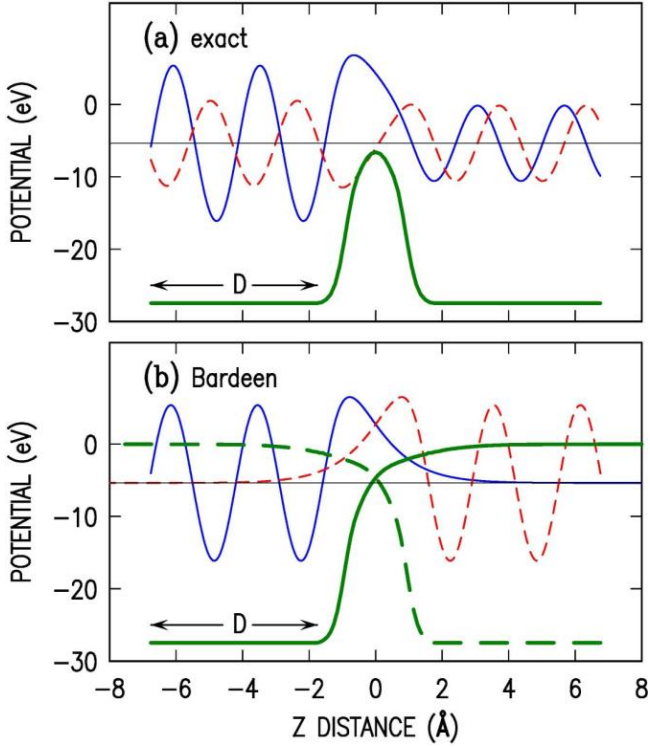

**Fig**. **8**. Model potentials (thick green solid and dashed lines) for two one-dimensional problems that are solved in order to investigate the accuracy of the Bardeen method. Wavefunctions (solid blue and dashed red lines) are shown for a state with energy 5.38 eV below the vacuum level (shown at 0 eV in the plots). (a) Potential from 2-ML WSe$_2$, truncated in the electrodes. The wavefunction is shown, with real part as solid blue line and imaginary part as dashed red line. (b) Potential from 1-ML WSe$_2$, truncated in the electrodes and shown for two separate electrodes on the left- and right-hand sides. Wavefunctions (purely real) for each electrode are shown.

Now we turn to a Bardeen solution for the problem, employing two electrodes each with potential obtained from the truncated 1-ML WSe$_2$ potentials. For a state of energy $E$ in each electrode, starting with a decaying exponential in the vacuum region with decay constant $\kappa = \sqrt{2m(E_{VAC} - E)/\hbar^2}$ for each electrode, the full wavefunctions are obtained by integrating Schrödinger's equation back through the barrier and into the electrode, and normalizing. The current for this state in one electrode is obtained by summing over a continuum of free-electron type states in the opposing electrode,

$$j_z^{\text{Bardeen}} = \frac{2\pi e}{\hbar} \sum_{k_j} |M|^2 \delta(E - E_j) \tag{11a}$$

$$= \frac{2\pi e}{\hbar} \frac{D}{2\pi} \int dE_j \frac{\sqrt{2m}}{\hbar\sqrt{E_j}} |M|^2 \delta(E - E_j) \tag{11b}$$

$$= \frac{2meD}{\hbar^3 k}|M|^2 \tag{11c}$$

with

$$M = \frac{\hbar^2}{2m}\left(\psi^* \frac{d\psi'}{dz} - \frac{d\psi^*}{dz}\psi'\right) \tag{12}$$

where $\psi$ denotes the wavefunction from one electrode and $\psi'$ is the wavefunction from the other, and the matrix-element $M$ is evaluated at the midpoint of the barrier. In Eq. (11b) we have inserted into the integrand the density-of-states for the continuum of free-electron type states, having wavevectors $k_j$ and energies $E_j$.

Figure 9 shows the ratio of the Bardeen to the exact solution, as a function of the energy of the state and for electrode thickness values of $3 - 5$ Å. This thickness enters both the exact and the Bardeen solutions through the normalization of the wavefunctions. We emphasize that we have not applied any specific boundary conditions on the far left of the left-hand electrode nor the far right of the right-hand one. Rather, our goal here is simply to compare the Bardeen to exact solutions as they pertain to the relatively small tunneling barriers that occur between the WSe$_2$ layers, i.e. arising from the potential within the van der Waals gap separating the electrodes. We find that for $D$ values of $3 - 5$ Å, the current obtained from Bardeen solution is typically a factor of $2 - 3$ times *smaller* than that from the exact current. The energies displayed in Fig. 9 cover the full range of values applicable to computations in the main body of this work: These energies within our one-dimensional model correspond to the *perpendicular component* of the energy, $E_\perp \equiv E_{3D} - \hbar^2|\mathbf{k}|^2/2m$, for an energy $E_{3D}$ of a state in a three-dimensional computation. Typical VB edges for the TMD materials lie at about $-5$ eV below the vacuum level, which for tunneling from Γ-point states then have the same values for $E_\perp$. However, for tunneling through a zone-edge (e.g. K-point state), we have $E_\perp \approx E_{3D} - 6$ eV, as discussed above in connection with Fig. 7. For CB edge states at the Γ-point, they will lie as high as about $-4$ eV below the vacuum level, i.e. at the upper end of the plotted curves of Fig. 9, and zone-edge CB states will lie about 6 eV below that. Certain special states can lie even lower on this energy scale, e.g. if the amplitude of their Fourier component within the first BZ is zero, so that their **k** value to be used in computing $E_\perp$ lies outside the first BZ (such states then decay correspondingly faster in the vacuum).

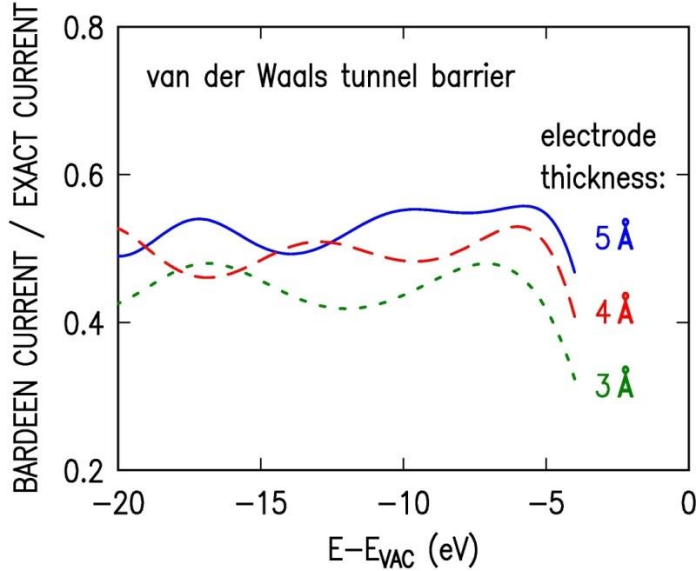

**Fig. 9**. Ratio of tunnel current from a Bardeen solution to that from an exact solution, for the model potentials shown in Fig. 8, as a function of the energy of the one-dimensional state relative to the vacuum level. Results are shown for three electrode thicknesses, *D*.

REFERENCES

[1] A. M. Ionescu and H. Riel, *Nature* 479, 329 (2011).
[2] M. Li, D. Esseni, G. Snider, D. Jena, and H. G. Xing, *J. Appl. Phys.* 115, 074508 (2015).
[3] M. Li, D. Esseni, J. J. Nahas, D. Jena, and H. G. Xing, *J. Elec. Dev. Soc.* 3, 200 (2015).
[4] D. Sarkar, X. Xie, W. Liu, W. Cao, J. Kang, Y. Gong, S. Kraemer, P. M. Ajayan, and K. Banerjee, *Nature* 526, 91 (2015).
[5] Y.-C. Lin, R. K. Ghosh, R. Addou, N. Lu, S. M. Eichfeld, H. Zhu, M.-Y. Li, X. Peng, M. J. Kim, L.-J. Li, R. M. Wallace, S. Datta, and J. A. Robinson, *Nat. Comm.* 6, 7311, (2015).
[6] R. Yan, S. Fathipour, Y. Han, B. Song, S. Xiao, M. Li, N. Ma, V. Protasenko, D. A. Muller, D. Jena, and H. G. Xing, *Nano Lett.* 15, 5791 (2015).
[7] G. C. Constantinescu and N. D. M. Hine, *Nano Lett.* 16, 2586 (2016).
[8] R. M. Feenstra, D. Jena, and G. Gu, *J. Appl. Phys.* 111, 043711 (2012).
[9] L. Britnell, R. V. Gorbachev, A. K. Geim, L. A. Ponomarenko, A. Mishchenko, M. T. Greenaway, T. M. Fromhold, K. S. Novoselov, and L. Eaves, *Nat. Comm.* 4, 1794 (2013).
[10] P. Zhao, R. M. Feenstra, G. Gu, and D. Jena, *IEEE Trans. Elec. Dev.* 60, 951 (2013).
[11] S. C. de la Barrera, Q. Gao, and R. M. Feenstra, *J. Vac. Sci. Technol. B* 32, 04E101 (2014).
[12] B. Fallahazad, K. Lee, S. Kang, J. Xue, S. Larentis, C. Corbet, K. Kim, H. C. Movva, T. Taniguchi, K. Watanabe, L. F. Register, S. K. Banerjee, and E. Tutuc, *Nano Lett.* 15, 428 (2014).
[13] S. C. de la Barrera and R. M. Feenstra, *Appl. Phys. Lett.* 106, 093115 (2015).
[14] P. M. Campbell, A. Tarasov, C. A. Joiner, W. J. Ready, and E. M. Vogel, *ACS Nano* 9, 5000 (2015).
[15] J. Bardeen, *Phys. Rev. Lett.* 6, 57 (1961).
[16] C. B. Duke, *Tunneling in Solids*, Solid State Physics, Suppl. 10 (New York: Academic, 1969).
[17] A. J. Bennett, C. B. Duke, and S. D. Silverstein, *Phys. Rev.* 176, 969 (1968).
[18] G. Kresse and J. Fürthmuller, *Phys. Rev. B* 54, 11169 (1996).
[19] G. Kresse, and D. Joubert, *Phys. Rev. B* 59, 1758 (1999).
[20] J. P. Perdew, K. Burke and M. Ernzerhof, *Phys. Rev. Lett.* 27, 3865 (1996).
[21] H. J. Monkhorst and J. D. Pack, *Phys. Rev. B* 13, 5188 (1976).
[22] C. Gong, H. Zhang, W. Wang, L. Colombo, R. M. Wallace, and K. Cho, *Appl. Phys. Lett.* 103, 053513 (2013).
[23] H. Liu, A. T. Neal, Z. Zhu, Z. Luo, X. Xu, D. Tománek, and P. D. Ye, *ACS Nano* 8, 4033 (2014).
[24] J. Cao, M. Pala, A. Cresti, and D. Esseni, In "2015 Joint International EUROSOI Workshop and International Conference on Ultimate Integration on Silicon" (EUROSOI-ULIS), pp. 245-248, IEEE, 2015.
[25] Á. Szabó, S. J. Koester, and M. Luisier, *IEEE Electron Device Lett.* 36, 514 (2015).
[26] F. Chen, H. Ilatikhameneh, Y. Tan, D. Valencia, G. Klimeck and R. Rahman, *arXiv preprint arXiv: 1608.05057* (2016).
[27] D. T. Do, S. D. Mahanti, and C. W. Lai, *Sci. Rep.* 5:17044 (2015).
[28] V. Tran, R. Soklaski, Y. Liang, and L. Yang, *Phys. Rev. B* 89, 235319 (2014).
[29] H. Lu and A. Seabaugh, *IEEE J. Electron Devices Soc.* 2, 44 (2014).

[30] G. Fiori, F. Bonaccorso, G. Iannaccone, T. Palacios, D. Neumaier, A. Seabaugh, S. K. Banerjee, and L. Colombo, *Nat. Nanotech.* 9, 768 (2014).

[31] R. Perricone, X. S. Hu, J. Nahas, and M. Niemier, In "2016 Design, Automation & Test in Europe Conference & Exhibition" (DATE), pp. 13-18, IEEE, 2016.

[32] J. Li, Y. Nie, K. Cho, and R. M. Feenstra, to be published.

[33] L. Britnell, R. V. Gorbachev, R. Jalil, B. D. Belle, F. Schedin, M. I. Katsnelson, L. Eaves, S. V. Morozov, A. S. Mayorov, N. M. Peres, A. H. Castro Neto, J. Leist, A. K. Geim, L. A. Ponomarenko, and K. S. Novoselov, *Nano Lett.* 12, 1707 (2012).

[34] (2011) International Technology Roadmap for Semiconductors. [Online]. Available: http://www.itrs.net/

[35] D. E. Nikonov and I. A. Young, *Proc. IEEE* 101, 2498 (2013).

[36] J. A. Stroscio, R. M. Feenstra, and A. P. Fein, *Phys. Rev. Lett.* 57, 2579 (1986).

[37] J. Tersoff and D. R. Hamann, *Phys. Rev. B* 31, 805 (1985).

[38] L. Britnell, R. V. Gorbachev, R. Jalil, B. D. Belle, F. Schedin, A. Mishchenko, T. Georgiou, M. I. Katsnelson, L. Eaves, S. V. Morozov, N. M. Peres, J. Leist, A. K. Geim, K. S. Novoselov, and L. A. Ponomarenko, *Science* 335, 947 (2012).

[39] R. M. Feenstra, Y. Dong, M. P. Semtsiv, and W. T. Masselink, *Nanotechnology* 18, 044015 (2007).